\documentclass[aps,pra,reprint,superscriptaddress,showpacs]{revtex4-1}
\usepackage[utf8]{inputenc}
\usepackage{color}
\usepackage{graphicx}
\usepackage{amssymb,amsmath,amsthm,amsfonts}
\usepackage{mathtools}

\usepackage{color}

\newcommand{\ming}[1]{\textbf{\color{red}[Ming: #1]}}

\newcommand{\comment}[1]{\textbf{\color{cyan}}}

\begin{document}
\title{Two-qubit gates in a trapped-ion quantum computer by engineering motional modes}
\author{Ming Li}
\affiliation{IonQ, College Park, MD 20740, USA}
\author{Jason Amini}
\affiliation{IonQ, College Park, MD 20740, USA}
\author{Yunseong Nam}
\affiliation{IonQ, College Park, MD 20740, USA}
\date{\today}

\begin{abstract}
A global race towards developing a gate-based, universal quantum computer that one day promises to unlock the never before seen computational power has begun and the biggest challenge in achieving this goal arguably is the quality implementation of a two-qubit gate. In a trapped-ion quantum computer, one of the leading quantum computational platforms, a two-qubit gate is typically implemented by modulating the individual addressing beams that illuminate the two target ions, which, together with others, form a linear chain. The required modulation, expectedly so, becomes increasingly more complex, especially as the quantum computer becomes larger and runs faster, complicating the control hardware design. Here, we develop a simple method to essentially remove the pulse-modulation complexity at the cost of engineering the normal modes of the ion chain. We demonstrate that the required mode engineering is possible for a three ion chain, even with a trapped-ion quantum computational system built and optimized for a completely different mode of operations. This indicates that a system, if manufactured to target specifically for the mode-engineering based two-qubit gates, would readily be able to implement the gates without significant additional effort.
\end{abstract}

\maketitle

\section{Introduction}

Quantum computational hardware is advancing fast, 
with much interest and investment from across the globe
\cite{Investment}. 
Two platforms, namely, superconducting and trapped-ion 
quantum computers (TIQC), are in the leads today, 
available for a commercial use~\cite{IBM,Rigetti,IonQ,Honeywell}. 
Unfortunately, these quantum computers are limited
in their computational power, mainly due to their
yet-to-mature underlying technologies that are used
to implement quantum gates. A two-qubit gate, being 
more sophisticated over a single-qubit gate in its
level of controls required, has traditionally been
and is expected to continue to be the bottleneck
in bettering the quality and increasing the power of
quantum computational systems.

A typical TIQC, with all the control hardware required,
is of substantial size comparable to an adult elephant.
When building a TIQC that operates on a laser-based gate,
a significant portion of the backroom consists of 
complicated control equipment, such as optics
and electronics, for modulating optical pulses, 
akin to the early days of classical computers
where the control hardware occupied a large physical volume of space.
If we are to learn from the
history of classical computer development,
to accelerate the TIQC hardware development,
there is likely a need to explore different technology options 
to realize quantum gate operations on a TIQC by shifting
the technical complication in certain less mature
domains of the control technology onto a more accessible, 
potentially new technology. By solving the backroom-size problem
and streamlining the engineering effort, an
eventual miniaturization of quantum computers may also be enabled.

In this paper, we take a first step towards addressing
this problem by trading the 
complexity in modulating the optical pulses for the 
complexity in implementing static trapping potentials. 
Specifically, we explore a possibility to greatly simplify 
pulse modulation for a TIQC at the expense of more complex 
trap voltage control and potentially more sophisticated 
trap design and fabrication. 
By demonstrating that our trade-off can readily be implemented
on today's system designed and built for a completely different mode
of operations on a three-ion chain, we make a concrete progress
towards a successful solution to the problem, instilling 
confidence that a co-design based approach between
the proof-of-principle theory work we show here 
and the prospective hardware development could
potentially advance the TIQC technology in the desirable direction.

\section{Traditional two-qubit gates on a trapped-ion quantum computer}

In this section, we briefly review a two-qubit XX gate,
typically used in a trapped-ion quantum computer.
Specifically, we assume a hardware configuration
of a linear chain of ions, each of which can be
addressed with individually addressing beam(s).

An XX gate induces quantum entanglement between 
two trapped-ion qubits $i$ and $j$, 
defined by the unitary operator
\begin{equation}
    {\rm XX}\left(\theta_{ij}\right) = e^{-i\theta_{ij}\left(
      \sigma_x^i\otimes\sigma_x^j\right)/2} \;, 
\label{eq:XX}
\end{equation}
where $\theta_{ij}$ denotes the degree of entanglement and
$\sigma_x^{i,j}$ are the Pauli-$x$ matrices acting on the qubit space $i$ or
$j$. To implement such a gate, the standard de-facto approach is
to use the widely adopted M{\o}lmer-S{\o}rensen (MS)
\cite{MS-1, MS-2} protocol, which exploits the coupling between the
individual qubit space and the shared motional space.
Ideally, the coupling is invoked during the gate operation only,
and the two spaces are decoupled from each other at the end of the gate.
More concretely, this means that, for an error budget of $\epsilon$,
we require~\cite{GREENPAP}
\begin{align}
    \alpha :=& \frac{4}{5} \sum_{p=1}^N \coth\frac{\hbar\omega_p}{2k_BT_p}
    ({\eta_p^i}^2+{\eta_p^j}^2)\left|\int_0^\tau g(t)
    e^{i(\omega_p t + \phi_p)} {\rm d}t\right|^2 \nonumber\\
    <& \varepsilon \;,
\label{eq:alpha}
\end{align}
where, for a linear chain with $N$ modes coupled with
the beams, $\alpha$ denotes
the residual coupling between motional modes $p$ and ions 
$i$ and $j$ after the gate operation,
$g(t)$ is the shape of the pulse that illuminates individual ions,
$\eta_p^{i,j}$ are the Lamb-Dicke parameters,
$\tau$ is the total gate time,
$\omega_p$ is the angular mode frequency, 
$\phi_p$ is the initial phase associated with mode $p$ at time $t=0$, 
$T_p$ is the temperature of the $p$-th mode,
$\hbar$ is the reduced Planck constant, and
$k_B$ is the Boltzmann constant.
Satisfying (\ref{eq:alpha}) 
ensures the contribution to the infidelity of the 
gate operation from the residual mode coupling is smaller
than $\varepsilon$.

As mentioned earlier, an XX gate defined in (\ref{eq:XX})  
has a parameter called the degree of entanglement $\theta_{ij}$. 
A formula for $\theta_{ij}$ is given by 
the sum of time-ordered double integrals, i.e.,
\begin{align}
    \chi_{ij} :=& \sum_{p=1}^{N} \eta_p^i\eta_p^j
    \int_0^\tau {\rm d}t_2 \int_0^{t_2} {\rm d}t_1
    g(t_2) g(t_1) \sin\left[\omega_p(t_2-t_1)\right] \nonumber\\
    =& \theta/4 \;.
\label{eq:chi}
\end{align}
A fully entangling XX gate requires $\theta_{ij} = \pi/2$, 
or $\chi_{ij} = \pi/8$.
An inexact implementation that induces uncertainty in $\chi_{ij}$
incurs quantum computational error.

To meet these two constraints (\ref{eq:alpha}) and (\ref{eq:chi})
and implement a high-fidelity two-qubit gate on a TIQC,
a host of sophisticated schemes have 
been developed.
For instance, given a mode-frequency spectrum, 
the pulse function $g(t)$ is modulated according to
the amplitude modulation (AM)~\cite{AM}, 
the frequency modulation (FM)~\cite{FM}, 
the phase modulation (PM)~\cite{PM}, 
and the power optimal AMFM~\cite{AMFM} schemes.
An important requirement to all these schemes is that 
one needs to ensure that the time-dependent electromagnetic (EM) 
wave seen by the ions follows the designed pulse shape faithfully.
As the pulse shape to be implemented becomes increasingly complex, 
which is typically the case as we add more qubits for more powerful
quantum computation or decrease the gate duration for faster quantum
computation, it becomes challenging to implement these schemes in practice.
Consider pulse-modulation hardware design. 
Sampling rate, frequency and amplitude range, rise and fall time, 
and signal distortion can start to limit the accuracy of the modulation.
The physical implementation of the modulation itself
via, e.g., an acousto-optic modulator (AOM) may pose yet another
challenge, as the input-output
response function might not be ideal, thus hindering
the accuracy of the EM wave seen by the ions.

\section{Motional mode engineering for simple two-qubit gate pulse}

In this section, we explore the idea of engineering the
motional modes themselves to help simplify the pulse shape $g(t)$.
Note the pulse shaping schemes discussed in the previous section
assumes the motional modes as a fixed input parameter.
Here we trade the pulse-shape complexity for 
overhead in engineering the mode spectrum. 
We show that, in return for a slightly more complex
static DC trapping field that, combined with the RF drive,
enables desirable control over the motional mode frequency
spectrum, we can essentially eliminate the need for 
pulse shaping.

\subsection{Gate requirement}
Recall that, to implement a two-qubit gate,
we need to satisfy (\ref{eq:alpha}) and (\ref{eq:chi}).
In our mode-engineering based approach,
we require the mode frequency spectrum to obey
\begin{description}
 \item[Condition 1] $\omega_p\tau/4 = k_p\pi$ where $k_p$ are
 positive integers.
\label{condition1}
\end{description}
We further require that the pulse $g(t)$ obey
\begin{description}
 \item[Condition 2] $g(t+\tau/2) = -g(t)$ for $t \in [0, \tau/2]$.
\label{condition2}
\end{description}
Such a pulse can be viewed as two
consecutive pulses of the same shape but with the opposite
phases. 
Using these conditions, while allowing for an error
$|\delta k_p| \ll 1$ in mode frequency 
$\omega_p = 4(k_p + \delta k_p)\pi/\tau$, 
we can rewrite the integral in (\ref{eq:alpha}) as
\begin{widetext}
\begin{align}
    \int_0^\tau g(t)e^{i(\omega_p t + \phi_p)} {\rm d}t = &
     \int_0^{\tau/2} g(t)e^{i(\omega_p t + \phi_p)} {\rm d}t 
     + \int_{\tau/2}^\tau g(t)e^{i(\omega_p t + \phi_p)} {\rm d}t \nonumber\\
    = &
     \int_0^{\tau/2} g(t)e^{i(\omega_p t + \phi_p)} {\rm d}t
     + \int_{\tau/2}^\tau g(t)e^{i[\omega_p (t-\tau/2) 
     + \omega_p\tau/2 + \phi_p]} {\rm d}t \nonumber\\
    = &
     \int_0^{\tau/2} g(t)e^{i(\omega_p t + \phi_p)} {\rm d}t
     - \int_0^{\tau/2} g(t)e^{i(\omega_p t + \phi_p 
     + 2\delta k_p\pi)} {\rm d}t \nonumber\\
    = & (1-e^{2i\delta k_p\pi}) 
     \int_0^{\tau/2} g(t)e^{i(\omega_p t + \phi_p)} {\rm d}t \;.
\label{eq:integral}
\end{align}
\end{widetext}
As $\delta k_p \to 0$, Eq.~(\ref{eq:integral}) vanishes, 
upperbounded by $\mathcal{O}[\max(|\delta k_p|)]$.
Therefore, as long as $\chi_{ij}$ in (\ref{eq:chi}) evaluates to 
a non-zero value, we can always scale $g(t)$ to satisfy the
gate requirements (\ref{eq:alpha}) and (\ref{eq:chi}),
assuming $\delta k_p\to 0$. 
In principle, \textbf{Conditions 1} and \textbf{2} are all that
we need to implement an XX gate.

\subsection{Pulse simplification and power optimization}
To further streamline the pulse shape requirement
while also lowering the power requirement,
we propose the following simple procedure.
We require the pulse to further satisfy
\begin{description}
 \item[Condition 3] $g(t) = \Omega\sin\left(\frac{2l\pi t}{\tau}\right)$
     for $t \in [0, \tau/2]$ where $l$ is a positive integer.
\end{description}
This choice simplifies the generation of the optical signal,
as the pulse is continuous throughout the gate duration
and the initial and final signal strength at $t = 0$ and $\tau$ 
is zero.
Combining now \textbf{Conditions 2} and \textbf{3}, we notice that 
we can separate these pulses into the groups with odd or even $l$.
For the pulses with odd $l$, the signal seen by the ion at
the half point of the pulse is differentiable and while
the pulses with even $l$ is not.
We can derive an analytic expression for $\chi_{ij}$ given by
\begin{widetext}
\begin{align}
    \chi_{ij} = 
    \begin{dcases*}
     \frac{\Omega^2\tau^2}{2\pi}
      \sum_{p=1}^{N} \eta_p^i\eta_p^j \frac{k_p}{4k_p^2 - l^2} \;,
       & \text{if $\nexists \; p=1,\cdots,N$ that $k_p = l/2$,} \\
     \frac{\Omega^2\tau^2}{2\pi} \left(
      \eta_{p_l}^i\eta_{p_l}^j\frac{3}{8 l} + 
      \sum_{\substack{p=1\\p\ne p_l}}^{N} 
       \eta_p^i\eta_p^j \frac{k_p}{4k_p^2 - l^2}
      \right)  \;,
       & \text{if $\exists \; p=1,\cdots,N$ that $k_p = l/2$.}
    \end{dcases*}
\label{eq:chi_analytic}
\end{align}
\end{widetext}
Here $p_l$ satisfies $2k_{p_l} = l$.
For a given set of Lamb-Dicke parameters and mode
frequencies, it is straightforward to pick an $l$ that requires the lowest
$\Omega$ to satisfy (\ref{eq:chi}), which corresponds
to the lowest peak-power requirement. To put it succinctly, we require
\begin{description}
 \item[Condition 4] $l$ is chosen so that the dimensionless 
 quantity $|\chi_{ij}/\Omega^2\tau^2|$
 is maximized according to (\ref{eq:chi_analytic}).
\end{description}

\subsection{Infidelity analysis}

Note there are two ways in which quantum computational errors may manifest,
non-zero $\alpha$ or inexact $\chi_{ij}$. 
Comparing to the underiable residual coupling
to the motional space, quantified by $\alpha$, which is difficult
to mitigate after the gate, error in the entanglement
angle quantified by $\chi_{ij}$ is more amenable to deal with.
For instance, static errors in $\chi_{ij}$ due to imprecise mode engineering 
or inaccurate knowledge of $\eta_p^{i,j}$ can readily
be calibrated away by using a slightly adjusted $\Omega$.
Inexact $\chi_{ij}$ induced by slowly drifting, temporal noise 
on the mode frequencies or Lamb-Dicke parameters can be mitigated 
by broadband composite pulse sequences designed for a two-qubit space~\cite{Murphy}.
Thus, in the following we focus on the accuracy needed for mode engineering, 
{\it i.e.}  the limit for $|\delta k_p|$, for a given error budget $\varepsilon$
on $\alpha$. 

To start, assuming non-zero $\delta k_p$, we write
\begin{widetext}
\begin{align}
    \alpha = 
    \begin{dcases*}
     \frac{l^2\Omega^2\tau^2}{5\pi^2}
     \sum_{p=1}^N \left[\coth\left(\frac{\hbar\omega_p}{2k_BT_p}\right)\right]
     ({\eta_p^i}^2+{\eta_p^j}^2)\left|
      \frac{e^{i\phi_p}(e^{4i\delta k_p\pi}-1)}{4(k_p+\delta k_p)^2 - l^2}
     \right|^2 \;,
     & \text{if $l$ is odd,} \\
    \frac{l^2\Omega^2\tau^2}{5\pi^2}
     \sum_{p=1}^N \left[\coth\left(\frac{\hbar\omega_p}{2k_BT_p}\right)\right]
     ({\eta_p^i}^2+{\eta_p^j}^2)\left|
      \frac{e^{i\phi_p}(e^{2i\delta k_p\pi}-1)^2}{4(k_p+\delta k_p)^2 - l^2}
     \right|^2 \;,
     & \text{if $l$ is even.}
    \end{dcases*}
\label{eq:alpha_precool}
\end{align}
\end{widetext}
Assuming the motional modes are sufficiently cooled to average phonon number $\bar{n}<1$,
which is readily achieved via sideband cooling schemes~\cite{SimCool}, we may let
$\coth\left(\frac{\hbar\omega_p}{2k_BT_p}\right) < 2$. Dropping now the modulus one term
$e^{i\phi_p}$, we can then bound $\alpha$ in (\ref{eq:alpha_precool}) by
\begin{widetext}
\begin{align}
    \alpha < 
    \begin{dcases*}
     \frac{2l^2\Omega^2\tau^2}{5\pi^2}
     \sum_{p=1}^N
     {(\eta_p^i}^2+{\eta_p^j}^2)\left|
      \frac{e^{4i\delta k_p\pi}-1}{4(k_p+\delta k_p)^2 - l^2}
     \right|^2 \;,
     & \text{if $l$ is odd,} \\
    \frac{2l^2\Omega^2\tau^2}{5\pi^2}
     \sum_{p=1}^N
     ({\eta_p^i}^2+{\eta_p^j}^2)\left|
      \frac{(e^{2i\delta k_p\pi}-1)^2}{4(k_p+\delta k_p)^2 - l^2}
     \right|^2 \;,
     & \text{if $l$ is even.}
    \end{dcases*}
\end{align}
\end{widetext}
Next, if we assume $\delta k_p$ is small, we can expand the right-hand
side to different orders of $\delta k_p$. This results in
\begin{widetext}
\begin{align}
    \alpha < 
    \begin{dcases*}
     \frac{32l^2\Omega^2\tau^2}{5}
     \sum_{p=1}^N
     ({\eta_p^i}^2+{\eta_p^j}^2)
     \frac{\delta k_p^2}{(4k_p^2 - l^2)^2}
     + \mathcal{O}(\delta k_p^3) \;,
     & \text{if $l$ is odd,} \\
    \alpha_0 + \frac{32l^2\Omega^2\tau^2\pi^2}{5}
     \sum_{\substack{p=1\\p\ne p_l}}^N
     ({\eta_p^i}^2+{\eta_p^j}^2)
     \frac{\delta k_p^4}{(4k_p^2 - l^2)^2}
     + \mathcal{O}(\delta k_p^5)\;,
     & \text{if $l$ is even,}
    \end{dcases*}
    \label{eq:alpha_analytic}
\end{align}
\end{widetext}
where
\begin{widetext}
\begin{align}
    \alpha_0 = 
    \begin{dcases*}
     0 \;,
       & \text{if $\nexists \; p=1,\cdots,N$ that $k_p = l/2$,} \\
     \frac{2\Omega^2\tau^2\pi^2}{5}({\eta_{p_l}^i}^2+{\eta_{p_l}^j}^2)
     \delta k_{p_l}^2 + \mathcal{O}(\delta k_p^3) \;,
       & \text{if $\exists \; p=1,\cdots,N$ that $k_p = l/2$.}
    \end{dcases*}
    \label{eq:alpha0}
\end{align}
\end{widetext}
We now insert $\alpha_0$ in (\ref{eq:alpha0}) 
into $\alpha$ in (\ref{eq:alpha_analytic}).
For a given error bound $\varepsilon$ for $\alpha$, 
we see that the required bound on $\delta k_p$ is
much less stringent for pulses with even $l$ than those with odd $l$,
if there is no $p$ such that $k_p = l/2$.
This suggests that it may be a good idea to indeed consider
this specific scenario, although in the remainder part of
the manuscript we chose to continue to consider both
even and odd $l$ pulses for completeness.
Considering that calibration or error-mitigation strategies are not
necessarily free, we caution that, in practice, 
it pays to carefully navigate the infidelity trade-off space.

An additional advantage of the proposed scheme is that 
the pulses we consider in our mode-engineering method
can readily be adapted to suppress crosstalk errors that arise from 
beam spill-over to nearby spectator ions. 
Note an $XX(\theta_{ij})$ gate, implemented according to
$(U_i \otimes U_j)^{-1} XX(\theta_{ij/2}) (U_i \otimes U_j) XX(\theta_{ij/2})$,
where $U_i^{-1} \sigma_x^i U_i = -\sigma_x$, suppresses the
aforementioned crosstalk errors to a second order.
Because of \textbf{Conditions 1} and \textbf{2}, the entanglement 
rotation angles accumulated in the first half of the gate
is already the same as accumulated in the second half, i.e.,
\begin{align}
 &\sum_{p=1}^{N} \eta_p^i\eta_p^j
 \int_0^{\tau/2}{\rm d}t_2 \int_0^{t_2} 
 {\rm d}t_1g(t_2)g(t_1) \sin\left[\omega_p(t_2-t_1)\right]
 \nonumber\\
 =&\sum_{p=1}^{N} \eta_p^i\eta_p^j
 \int_{\tau/2}^\tau {\rm d}t_2 \int_{\tau/2}^{t_2} 
 {\rm d}t_1g(t_2)g(t_1) \sin\left[\omega_p(t_2-t_1)\right]
 \nonumber\\
 =& \;\chi_{ij}/2 \;.
\end{align}
By flipping the pulse phase
on the qubit ions as in \textbf{Condition 2} and leaving
the phase of the spectator ions unchanged, we can immediately
implement the crosstalk suppression sequence.

\section{Example: 3-ion case}

\begin{figure}
    \centering
    \includegraphics[trim=80 0 0 0, clip, width=\columnwidth]{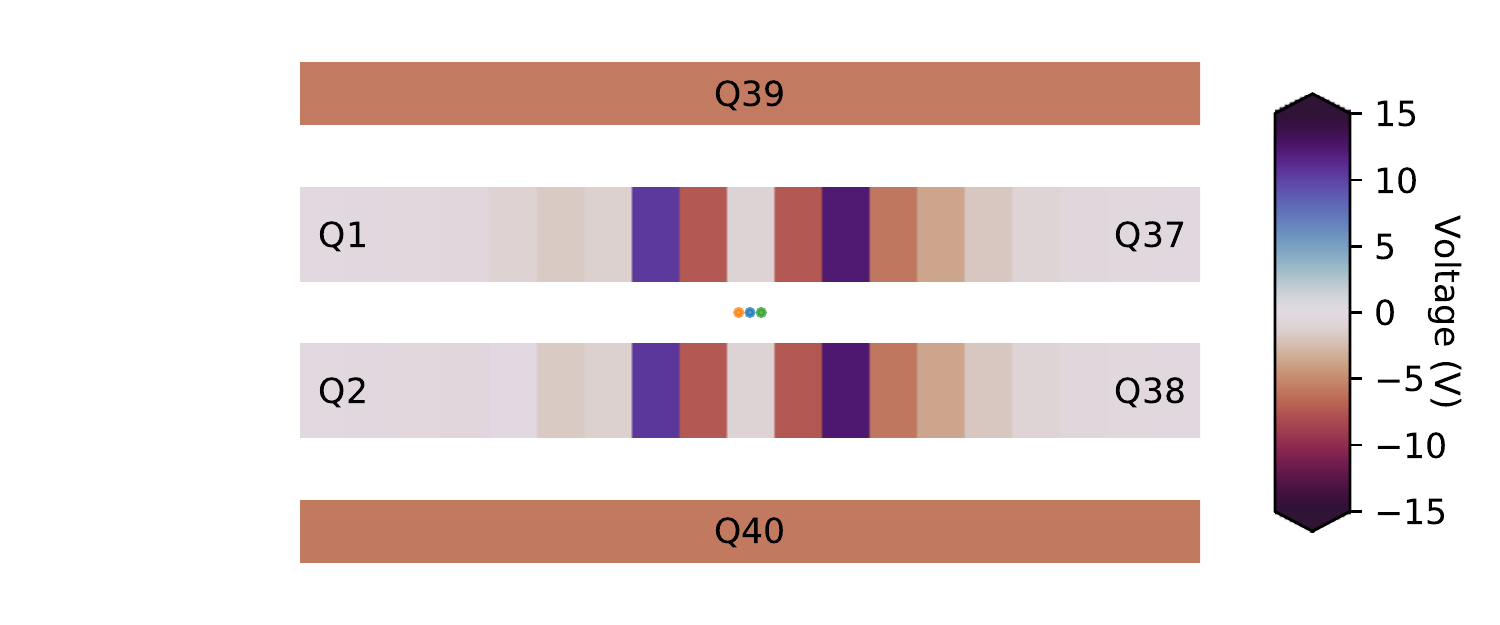}
    \caption{ Numerically solved
    trapping voltage color map for the 
    ``quantum'' zone control electrodes
    of the HOA 2 trap. The three blue dots
    in the middle indicates where the three
    ion chain is trapped. The figure is
    not to scale.}
    \label{fig:trap}
\end{figure}

\begin{figure}
    \centering
    \includegraphics[trim=10 0 0 0, clip, width=1.05\columnwidth]{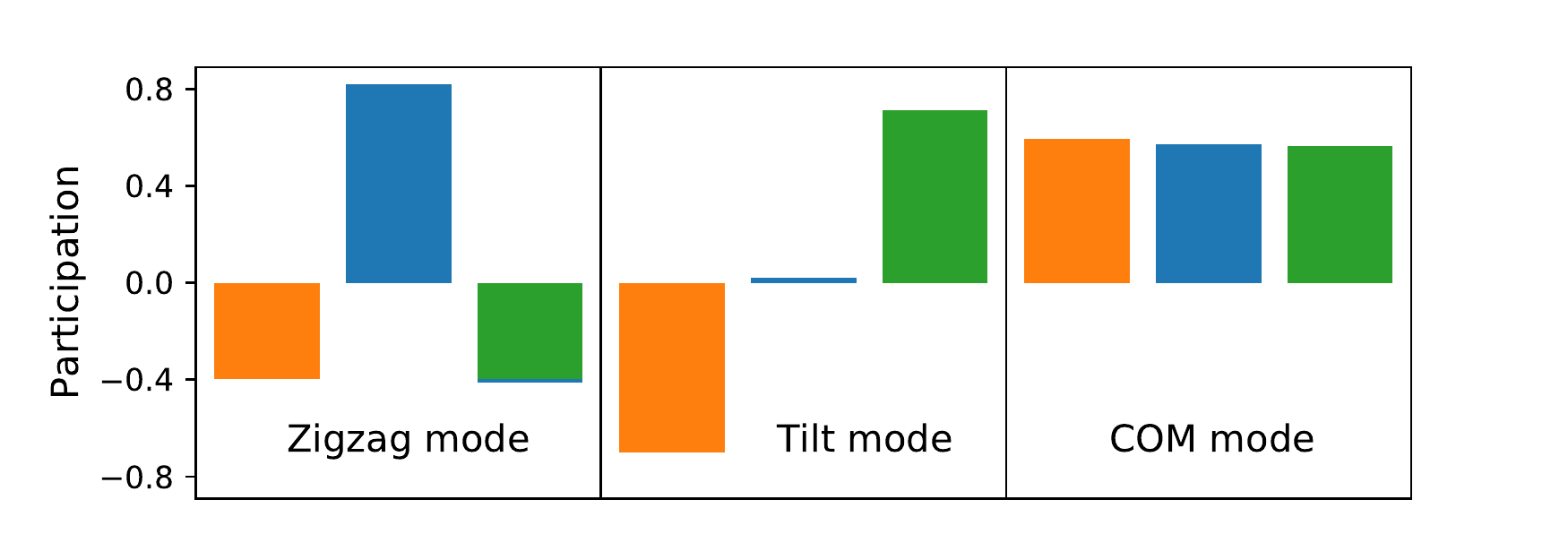}
    \caption{Mode participation for the three-ion chain.
    Color orange represents ion 0, blue represents ion1,
    and green represents ion 2. The zig-zag mode has the
    lowest mode frequency while the COM mode has the
    highest.}
    \label{fig:part}
\end{figure}

\begin{figure}
    \centering
    \includegraphics[width=1.05\columnwidth]{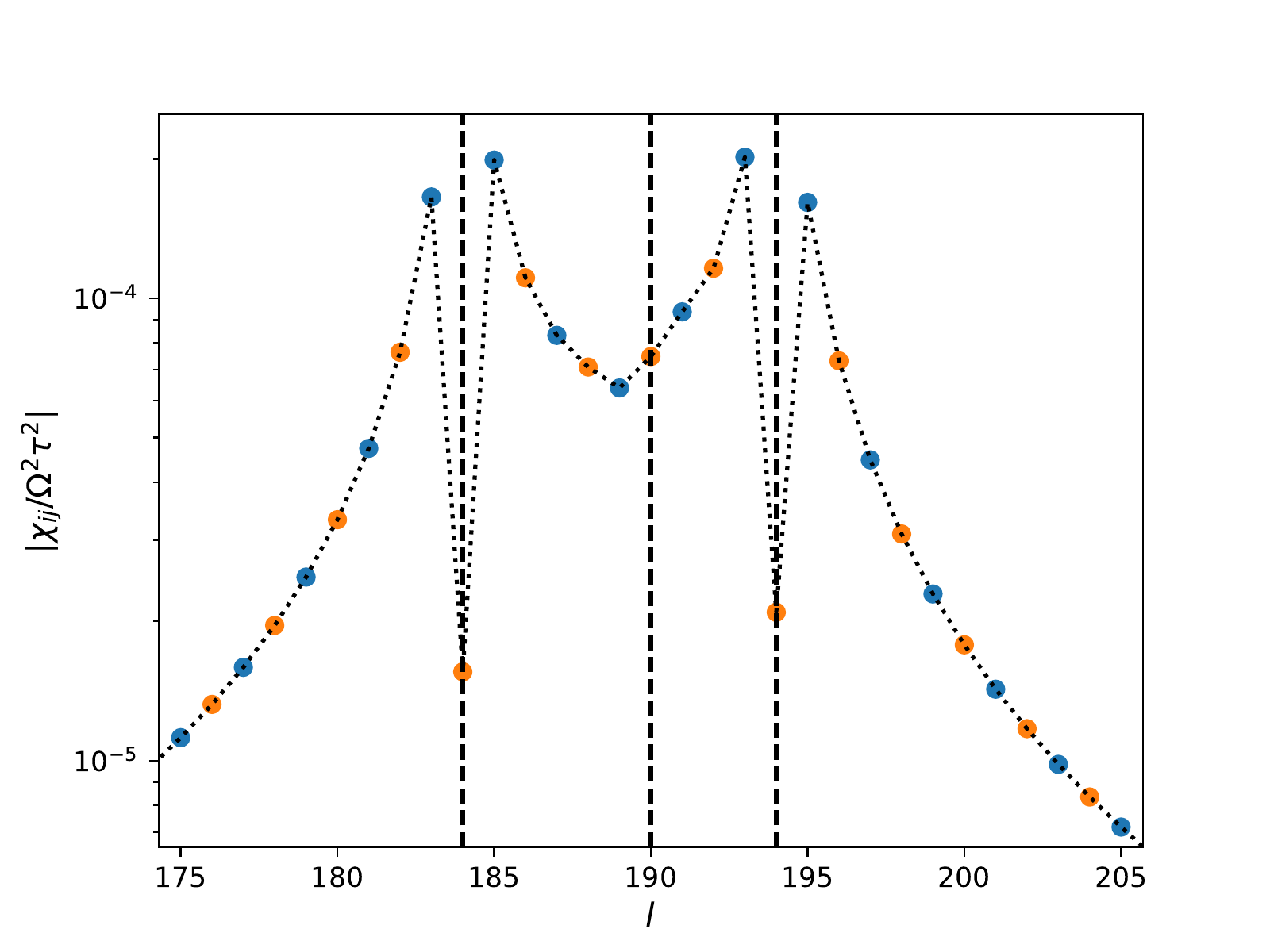} 
    \caption{$|\chi_{ij}/\Omega^2\tau^2|$ versus $l$
    in \textbf{Condition 4} for an XX-gate implementation
    between ion 0 and 1. The values for even $l$s are shown
    as filled orange circles and the ones for odd $l$s are
    shown as filled blue circles. The vertical black dashed-lines
    correspond to the $l = 2k_p$ of the ideal mode frequencies.
    We pick $l = 193$ as the odd $l$ pulse and $l = 192$
    as the even $l$ pulse for their lowest power requirement
    respectively.}
    \label{fig:mat}
\end{figure}

\begin{figure}
    \centering
    \includegraphics[width=1.05\columnwidth]{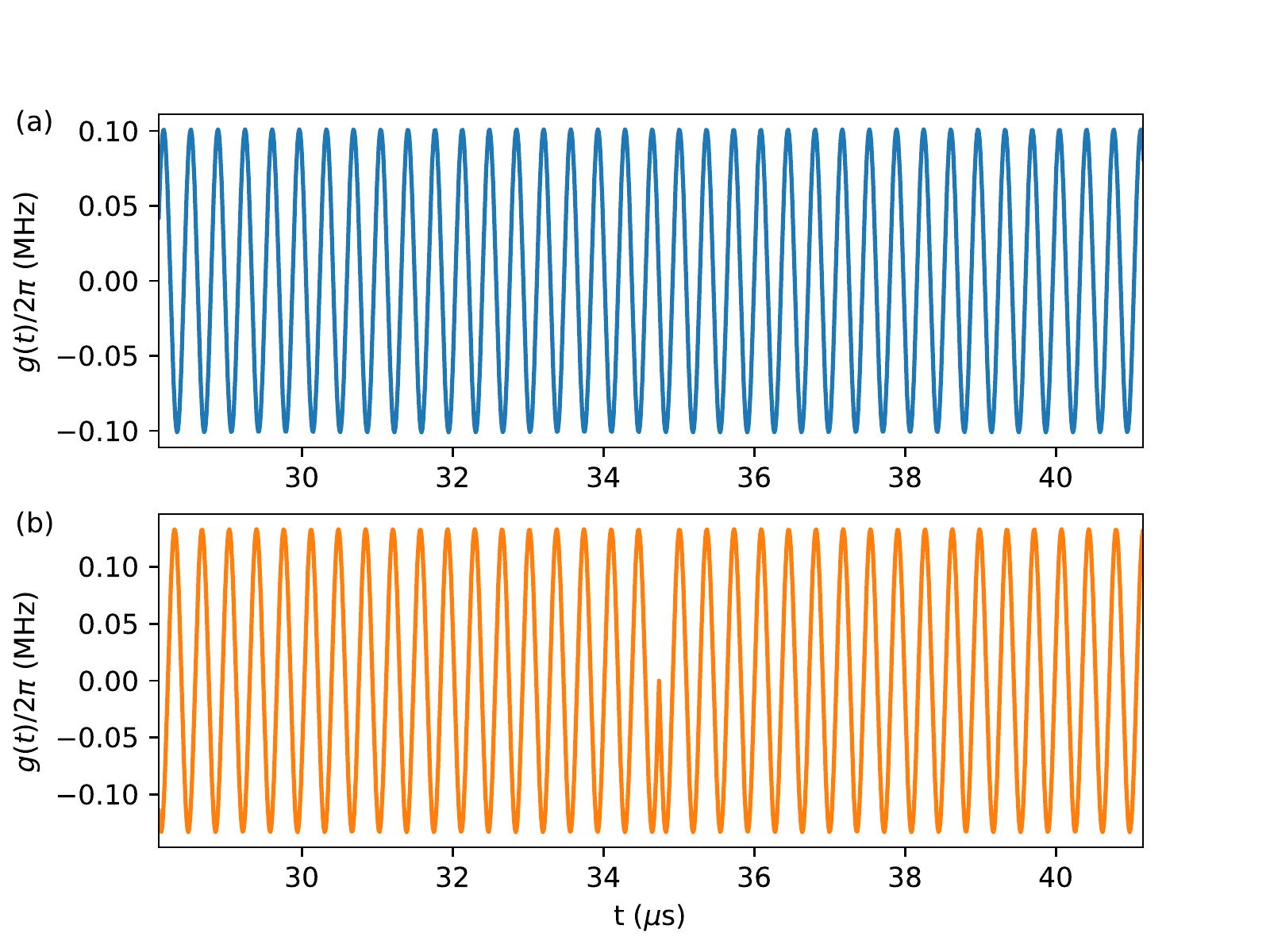} 
    \caption{Pulses of (a) odd, $l = 193$, and (b) even, 
    $l = 192$, with the lowest power requirement respectively
    for an XX-gate implementation between ion 0 and 1.
    The pulse with odd $l$ is infinitely
    differentiable at the half way point of the gate at
    $t = 34.733 \mu$s while the pulse with even $l$ has a 
    cusp.}
    \label{fig:pulses}
\end{figure}

\begin{figure}
    \centering
    \includegraphics[width=1.05\columnwidth]{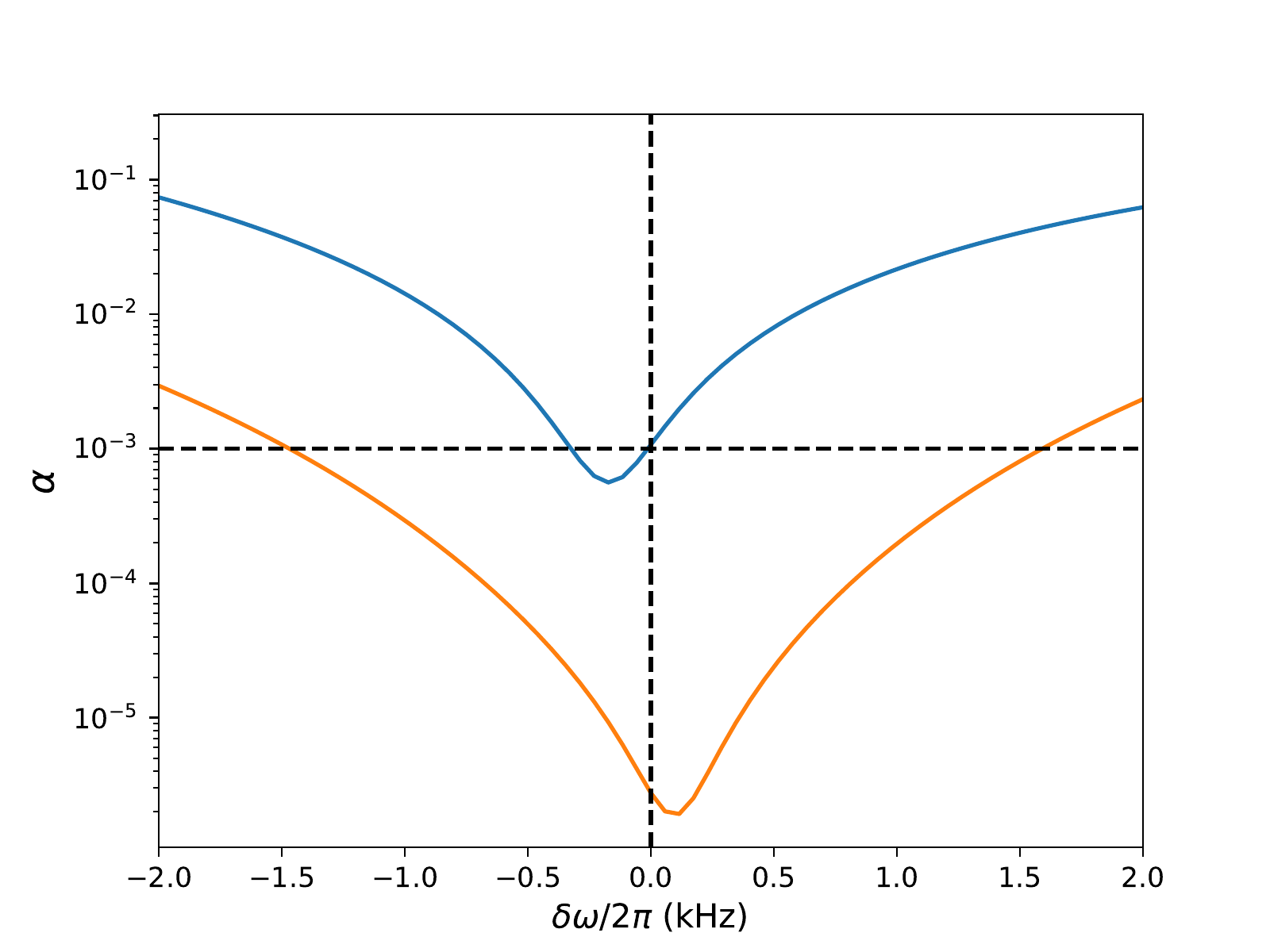}
    \caption{$\alpha$ as a function of common mode shift 
    $\delta\omega$ for the odd, $l = 193$, (blue) and 
    even, $l = 192$, (orange) pulses with the 
    highest powers respectively for an XX-gate implementation
    between ion 0 and 1. 
    }
    \label{fig:alpha}
\end{figure}

Here we demonstrate our mode-engineering method. 
Specifically in this section we consider a three-ion chain,
trapped in a Sandia High Optical Access (HOA) 2 trap
\cite{HOA} with individually addressing Raman
beams perpendicular to the chain which couple to
the three transverse modes. 

To start, due to limited voltage range available in the HOA 2 trap, 
it is difficult to generate a scalable, analytical electrode-voltage solution
that will result in a linear ion chain that satisfies \textbf{Condition 1} 
for arbitrarily many ions. 
For the three-ion case, however, we can numerically optimize.
Optimizing over the mode frequencies 
using possible DC controls provided by the HOA 2 trap
to fulfill \textbf{Condition 1}, while also maintaining
an ion spacing of $\sim 4.3\mu$m constrained by the geometry
of individually addressing Raman beams, one possible 
solution found was the gate time of $\tau = 69.466 \mu$s and $k_p = 92$, $95$,
and $97$ for the zigzag, the tilt, and the center-of-mass
(COM) modes, respectively.

Figure~\ref{fig:trap} shows the numerically solved
trapping voltage needed on the DC control electrodes in the
``quantum'' zone -- the zone where quantum gate operations
are performed -- of the HOA 2 trap. The voltage
configuration traps the linear three ion chain
with $\sim 4.3 \mu$m spacing about $71 \mu$m above
the trap surface, when combined with the RF drive
at $50.6$ MHz frequency. In order to produce
the desirable mode frequencies, we apply
$289.71$ V on the RF rails. The DC electrodes
generates a slightly higher radial squeeze on the center
ion than on the two end ions, which generate
the zigzag mode at $2.649 \times 2\pi$MHz, the tilt
mode at $2.735 \times 2\pi$ MHz, and the COM mode at 
$2.793 \times 2\pi$ MHz. These frequencies deviates slightly
from the ideal frequencies given by \textbf{Condition 1}
with the $k_p$ mentioned earlier. However, impact of the
error is small, which we describe in the later part of this
section. The mode participation vectors
$\nu_p^i$ are shown in Fig.~\ref{fig:part}, which are
related to the Lamb-Dicke parameters by
$\eta_p^i = \nu_p^i\sqrt{\hbar/2m\omega_p}$ with $m$ being
the mass of the ions.

Without loss of generality, we now focus on implementing 
an XX gate between ions zero and one.
We apply \textbf{Conditions 2} and \textbf{3} to
$g(t)$, which allows us to use (\ref{eq:chi_analytic})
to compute $|\chi_{ij}/\Omega^2\tau^2|$ for different
choices of $l$, as shown in Fig.~\ref{fig:mat}. Following
\textbf{Condition 4}, we pick $l = 192$ and $193$ for
their low power requirement on $\Omega$ for even and odd
$l$ pulses, respectively. The pulse with $l = 192$
requires slightly more power with $\Omega/2\pi = 0.133$ MHz
comparing to $0.101$ MHz for the pulse with $l = 193$.
The middle section of the pulses
are illustrated in Fig.~\ref{fig:pulses} to show the
cusp for $l = 192$ at the half-way point of the pulse,
which does not exist for $l = 193$. However, if we
opt to apply a single qubit gate rotation at that point
instead of changing the optical phase in order to 
suppress first order cross-talk error, then the
even $l$ pulse will not have the cusp which makes it
more amenable to better implement in practice.

Having determined the pulses, we next examine 
the stability of the fidelity
of the pulses with respect to the error in mode frequency
engineering and possible experimental noise. 
We numerically evaluate (\ref{eq:alpha}) using
non-ideal mode frequencies, discussed earlier in the
context of the limitations of the HOA 2 trap design,
plus a variable common mode frequency error $\delta\omega$ applied to
all three modes. The results are shown in Fig.~\ref{fig:alpha}.
The blue curve represents the pulse with $l = 193$
and the orange curve represents the pulse with $l = 192$.
At $\delta\omega = 0$, the infidelity due to residual
motional couplings as quantified by $\alpha$ solely
comes from the static error of mode engineering or in other
words the deviation of realistically generated frequencies
from the ideal ones. For $\delta\omega$ of non-zero values,
the $\alpha$s represent infidelities due to experimental
error or noises which causes the mode frequencies to from
the designed ones. As expected from (\ref{eq:alpha_analytic})
and (\ref{eq:alpha0}), the pulse with even $l$ has much
better infidelity compared to the pulse with odd $l$,
owing to the $\delta\omega^4$ dependence for even $l$ instead of 
$\delta\omega^2$ dependence for odd $l$. Of course, 
such an advantage exists when one chooses $l \neq 2k_p$ of any mode,
which is the case for our current example. 
The trade off is that $\sim 30\%$ higher power requirement
for $l=192$, compared to that for $l=193$. Considering
the additional power requirement may be prohibitive in 
terms of implementation, the choice of $l$ 
should be made on a case by case basis while fully taking the
hardware limit into consideration.

\section{Discussion}

\comment{
\ming{
For Jason:
\begin{itemize}
    \item Co-designing trapping waveform, traps,
    and individual addressing with possibly uneven spacing
    and long chains.
    \item RF stability, stray field compensation,
    and other electric field noises.
\end{itemize}
We could also discuss possibilities of using
optical trapping potentials to engineer modes.
}
}
\newcommand\vxi{\ensuremath{{\bf x}_i}}
\newcommand\vxj{\ensuremath{{\bf x}_j}}

So far in this paper we demonstrated a method to engineer
the motional modes to enable an implementation of a MS gate
and showed a viability of it using an explicit example
of a three-ion chain. In particular, we focused on a system
manufactured for a completely different MS gate implementation
architecture. This motivates us to consider if a co-design
effort may be made to better implement the mode-engineering
based approach in a scalable fashion. Below, we discuss this
point briefly with an eye toward the technical requirements
and challenges.

Note the radial and axial mode structures of a chain of ions 
in an ion trap can be controlled through a combination of 
the ion spacing and local potentials around each ion. 
To be more concrete, the local confining potential $U$ 
of a chain of ions $i\in[0,N)$ is
\begin{eqnarray}
U(\{\vxi\}) & = & 
\frac12 \sum_i \vxi^T P_{i} \vxi  
+ \frac 12  \sum_i \vxi^T A_{i} \vxi \nonumber \\
& & + \frac 12  \sum_{ij} \vxi^T C_{ij} \vxj
+ O({\bf x}^3),
\end{eqnarray}
where the displacements $\vxi$ are measured from the respective equilibrium locations of the ions, $A$ encodes the electrostatic potentials applied to the trapping electrodes, $C$ encodes the Coulomb interactions between pairs of ions and depends only on the equilibrium ion locations, and $P$ encodes the standard pseudopotential applied by the RF potentials. 
Up to $O({\bf x}^3)$ then, the potential can be succinctly expressed as 
$U \approx \vec{{\bf x}}^T K \vec{{\bf x}}$, where 
\begin{widetext}
\begin{eqnarray}
K & = P + A + C = & 
 \begin{bmatrix}
    P_0+A_0+C_{00} & C_{01} &  \cdots & C_{0(N-1)} \\
    C_{10} & P_1+A_1+C_{11} & \cdots & C_{1(N-1)} \\
    \vdots & \vdots & \ddots & \vdots \\
    C_{(N-1)0} & C_{(N-1)1} & \cdots & 
        P_{N-1}+A_{N-1}+C_{(N-1)(N-1)} \\
  \end{bmatrix}.
\end{eqnarray}
\end{widetext}

Note the mode frequencies are given by the eigenvalues of $K$.
Thus, we can control the mode structure by changing the matrix elements of $K$.
For instance, in most ion traps, $P$ is generated from a single source 
and can be scaled by changing the RF potential and/or frequency. 
For surface traps, and other traps with radial asymmetry, 
ion displacements from the RF null can further introduce changes to $P$. 
Multi-electrode traps surface traps, such as the Sandia HOA~\cite{HOA}, 
provide arrays of electrodes that allow one to tune $P$ and $A$ and, 
through adjustments to the equilibrium locations, $C$. 
Optical dipole potentials can further be used to modify $A$.
When designing a tailor-made system that operates on the basis of
the mode engineered MS gate, a careful consideration then must be
given to the trade-offs between ion placement, micromotion, 
spatial twisting of mode participation (the eigenvectors of $K$), and mode spacing.

\section{Summary and outlook}

When designing traps for particular ion configurations, 
one can optimize the electrodes structure to match the desired potentials.
With appropriate technological advancements, 
an increased number of trap electrodes and their shapes
may be leveraged to induce the desired potentials.
Further adjustments to the ion potentials could be made
using potentials induced from non-trap sources as well.
Whether these efforts are more amenable to higher quality
quantum computer manufacturing remains to be seen.
Our work provides one of the first, possible ways to redistribute 
the technical challenges that must be overcome to build a
practical quantum computer and will certainly not be the last.


\begin{thebibliography}{99}

\bibitem{Investment}
E. R. MacQuarrie, C. Simon, S. Simmons, and E. Maine, The emerging commercial landscape of quantum computing. {\it Nat. Rev. Phys,} {\bf 2}, 596–598 (2020).

\bibitem{IBM}
IBM Research. 
Quantum Experience. 
\href{http://www.research.ibm.com/quantum/}{http://www.research.ibm.com/quantum/},
Accessed November 16, (2020).

\bibitem{Rigetti}
Rigetti.
Amazon Bracket Hardware Provider.
\href{https://aws.amazon.com/braket/hardware-providers/rigetti/}{https://aws.amazon.com/braket/hardware-providers/rigetti/},
Accessed November 16, (2020).

\bibitem{IonQ}
\href{https://aws.amazon.com/braket/hardware-providers/ionq/}{https://aws.amazon.com/braket/hardware-providers/ionq/},
Accessed November 16, (2020).

\bibitem{Honeywell}
\href{https://www.honeywell.com/en-us/company/quantum/quantum-computer}{https://www.honeywell.com/en-us/company/quantum/quantum-computer},
Accessed November 16, (2020).

\bibitem{MS-1} 
K. M{\o}lmer, A. S{\o}rensen, 
Multiparticle Entanglement of Hot Trapped Ions, 
{\it Phys. Rev. Lett.} {\bf 82}, 1835-1838 (1999). 

\bibitem{MS-2} 
K. M{\o}lmer, A. S{\o}rensen, 
Entanglement and Quantum computation with ions 
in thermal motion,
{\it Phys. Rev. A} {\bf 62}, 022311 (2000). 

\bibitem{GREENPAP} Y. Wu, S.-T. Wang, L.-M. Duan, 
Noise Analysis for High-Fidelity Quantum Entangling Gates in an Anharmonic Linear Paul Trap,
{\it Phys. Rev. A} {\bf 97}, 062325 (2018). 

\bibitem{AM}
S.-L. Zhu, C. Monroe, L.-M. Duan,
Arbitrary-speed quantum gates within large ion crystals through minimum control of laser beams,
{\it Europhys. Lett.} {\bf 73}, 485 (2006).

\bibitem{FM}
P.~H. Leung, K.~A. Landsman, C. Figgatt, N.~M. Linke, C. Monroe, K.~R. Brown,
Robust 2-qubit gates in a linear ion crystal using a frequency-modulated driving force,
{\it Phys. Rev. Lett.} {\bf 120}, 020501 (2018).

\bibitem{PM}
T.~J. Green, M.~J. Biercuk, 
Phase-modulated decoupling and error suppression in qubit-oscillator systems,
{\it Phys. Rev. Lett.} {\bf 114}, 120502 (2015).

\bibitem{AMFM}
R. Blumel, N. Grzesiak, and Y. Nam,
Power-optimal, stabilized entangling gate 
between trapped-ion qubits,
\href{https://arxiv.org/abs/1905.09292}{https://arxiv.org/abs/1905.09292} (2019).

\bibitem{Murphy}
Daniel C. Murphy, Kenneth R. Brown,
Controlling error orientation to improve quantum algorithm success rates.
{\it Phys. Rev. A} {\bf 99}, 032318 (2019).

\bibitem{SimCool}
J.-S. Chen, K. Wright, N. C. Pisenti, D. Murphy, K. M. Beck, K. Landsman, J. M. Amini, and Y. Nam,
Efficient-sideband-cooling protocol for long trapped-ion chains,
{\it Phys. Rev. A} {\bf 102}, 043110 (2020).

\bibitem{HOA}
P. L. W. Maunz, Sandia National Laboratories Report
No. SAND2016-0796R (2016).

\end{thebibliography}
\end{document}